# An Enhanced Edge Adaptive Steganography Approach using Threshold Value for Region Selection

Sachin Mungmode[1], R. R. Sedamkar[2] and Niranjan Kulkarni[3]

[1] Department of Computer Engineering, Mumbai University, Mumbai, India
sachinmungmode30@gmail.com
[2] Department of Computer Engineering, Mumbai University, Mumbai, India
rr.sedamkar@thakureducation.org
[3] Department of Computer Engineering, Mumbai University, Mumbai, India
niranjan.kulkarni@thakureducation.org

*Abstract*

*This paper attempts to improve the quality and the modification rate of a Stego Image. The input image provided for estimating the quality of an image and the modified rate is a bitmap image. The threshold value is used as a parameter for selecting the high frequency pixels from the Cover Image. The data embedding process are performed on the pixels that are found with the help of Threshold value by using LSBMR. The quality of an image is estimated by the value of PSNR and the modification rate of an image is estimated by the value of MSE. The proposed approach achieves about 0.2 to 0.6 % of improvement in the quality of an image and about 4 to 10 % of improvement in the modification rate of an image compared to the edge detection techniques such as Sobel and Canny.*

## 1. Introduction

Steganography has a Greece rootage where the word 'Steganos' means 'Covered' and 'Graptos' means 'Writing' [1]. Steganography is also known as 'Secret Writing'. Steganography is a subdivision of information security that hides the secret and crucial information into an innocent medium in such a way that no one except recipient can extract the information [2]. The digital medium can be an image, audio or video file [3]. In innovative life, steganography is working for many purposes such as embedding copyright, watermark and covert communication. The Steganalysis process targets to unwrap the presence of hidden messages in those stego medium. The existence of steganalytic algorithms helps to find out the medium is a cover or not with a more probability than random guessing.

Following valuable properties must be taken into circumstances while implementing steganographic approaches.

1) *Imperceptibility:* Imperceptibility remains the firsthand destination of steganography. The image must look exactly alike before and after the modification. A person must not be able to see the changes in image.
2) *Embedding Capacity:* Capacity is the amount of information that can be embedded in a particular medium. Capacity seems to be an issue for steganography as the size of the medium and information to hide goes on changing. So, the embedding capacity remains a probatory factor in steganography.

3) *Robustness:* Robustness is the level of difficulty required to demolish embedded information without destructing the cover medium and extracting the required information.
4) *Undetectability:* Undetectability is the power of determining whether the cover medium contains embedded information or not by employing different technical means. Undetectability tries to hold back the fact that a message is being transferred from sender to receiver.

Image steganography is the subdivision of steganography where digital images are used as bearer file formats for information [4]. The Joint Photographic Experts Group Format (JPEG), Graphics Interchange Format (GIF), Bitmap Picture format (BMP) and Portable Network Graphics format (PNG) are the majority popular image file formats that are being shared on the internet.

Edge adaptive steganography is a peculiar case of the spatial or transform domain steganographic technique [5]. The edge adaptive technique is based on area of interest where edge detection techniques are used to choose a region randomly for data embedding. The edges of image are detected using edge detection techniques like Sobel and Canny. As it selects the region of high intensity in an image, the smooth areas stay unmodified. Advantage of this technique is the quality of an image and the modification rate of an image improves.

This paper contains the results of execution of an enhanced technique on various images. Following are the sections explained in the paper.

1) The first section of the paper gives the brief idea about steganography and related work on Steganography.
2) The second section deals with the objective and the scope of the result, which are delivered by execution of the algorithm.
3) The third section gives overview about Sobel edge detection technique, Canny edge detection technique and LSBMR algorithm.
4) The fourth Section describes proposed technique for data embedding and data extraction.
5) The fifth section shows experimental results drawn by executing the proposed techniques on nearly 100 images.

## 2. Related Research

The Least Significant Bit (LSB) is the basic technique used for steganography. LSB technique is used to hide out the secret message bits into the least significant bits of the cover image. The LSB is a swapping type of technique which swaps bits of cover image by secret message bit. Simple LSB steganography is easily perceptible by steganalytic methods like Regular Singular (RS) and Chi-square analysis [3].

LSBM (Least Significant Bit Matching) technique employed slight modification to the LSB method [6]. In LSBM, if the private bit does not match the LSB of the cover image pixels, the pixel values of an image are modified by adding one or subtracting one [7]. Statistically, the probability of increasing or decreasing of pixel value for an image is the same and the imbalance artifacts introduced in LSB substitution will be easily avoided. Therefore, the common approaches that are utilized to determine the presence of LSB replacement are totally useless at detecting the LSBM. The proposed method gives an option to either add or subtract one from the cover image pixel at random. LSBM provided distortion and resistance to the Steganalysis [8]. The drawback of LSBM method is that the detection of the existence of the hidden messages using the Centre of Mask component of the Histogram Characteristic Function [3] based detectors is less efficient.

Jarno Mielikainen proposed LSBMR (Least Significant Bit Matching Revisited) which uses grayscale cover images [9]. The data embedding is performed on the pair of pixels at a time. LSBMR eliminated the asymmetry property caused by basic LSB. The Pixel-Value Differencing (PVD) [10] technique uses different approach where the numbers of embedded bits are determined by the difference between a pixel and its neighbor [11] [12]. Edge Adaptive LSBMR (EALSBMR) proposed by Weiqi Luo, Fangjun Huang and Jiwu Huang.

Zohreh Fouroozesh and Jihad Al Jaam proposed Edge Adaptive Steganography using the Sobel edge detection technique [13]. The technique is employed on grayscale images and LSBMR is used for data embedding [1].

## 3. Objective of the Work

The Sobel and Canny edge detection techniques work well with gray scaled images. Therefore, the color image need to be converted to a grayscale image for the data embedding process and this image format conversion has a possibility of data loss. So, the improvised approach is carried out on color images with LSBMR as a data embedding algorithm. The proposed mechanism aims to improve the quality of an image and modification rate of stego image. PSNR and MSE are the parameters for evaluating the quality of an image and the modification rate.

## 4. Sobel Edge Detection

The Sobel edge detector algorithm uses Sobel's operator to extract edges [1]. For lower embedding rates the sharp edges are used for holding information and as the rates increase the algorithm adaptively is adjusted to use less sharp edges. Sobel operator is an orthogonal gradient operator. It detects edges of the point, according to its adjacent points. The Sobel edge operator performs a convolution in x-direction with the help of incline component called $G_x$ on selected image. The Sobel edge operator performs a convolution in y-direction with the help of incline component called $G_y$ on selected image. The incline components $G_x$ and $G_y$ are shown in figure 1.

| +1 | +2 | +1 |
|----|----|----|
| 0  | 0  | 0  |
| -1 | -2 | -1 |

$G_x$

| +1 | 0 | -1 |
|----|---|----|
| +2 | 0 | -2 |
| +1 | 0 | -1 |

$G_y$

Figure 1. Incline components $G_x$ and $G_y$ with their masks

These kernels are designed to react maximally to edges running in vertical and horizontal direction relative to the pixel grid, one kernel for each of the two perpendicular orientations [2]. The kernels are employed individually on input image to bring out individual measurements of incline components $G_x$ and $G_y$. The combination of these two gives the absolute magnitude of the incline at each point and the orientation of that incline. The incline magnitude is given by:

$$|G| = \sqrt{G_x^2 + G_y^2} \tag{1}$$

Absolute magnitude of the incline is given by-

$$|G| = |G_x| + |G_y| \tag{2}$$

The angle of orientation of the edge giving rise to the spatial incline is given by-

$$\theta = -\tan^{-1}(G_y/G_x) \tag{3}$$

*Advantages:*
- Provides a simple estimation to incline magnitude.
- Easy in detecting edges and their orientation.

*Disadvantages:*
- Sensitivity to the noise, edge detection and orientation of the edges.
- Inaccuracy to the incline magnitude.

# 5. Canny Edge Detection

The canny edge detector algorithm uses Gaussian operator for extracting edges of an image [6]. To lower embedding rate the sharp edges are used for holding information and as the rate increases the algorithm adaptively is adjusted to use less sharp edges [7]. The three criteria used by Canny for detecting edges are-
1) *Low error rate*- Every edge form the image should be detected and there should be no response for non-edges.
2) *Localization of edge pixels*- The edge points of the image should be well localized. That means, the distance between edge pixels as found by the detector and the actual edge is to be at minimum.
3) *One response to Single edge*- There should be single response to single edge. This will eliminate multiple responses to a single edge.

The Canny edge detection algorithm performs following steps in the edge detection process.
*Step 1:* Filtering of noise from the input image is carried out using Gaussian filter. Gaussian filter calculates suitable mask and then Gaussian smoothing is done using standard convolution methods.
*Step 2:* Find out the edge strength by taking the gradient of the image. The gradient of an input image is estimated using a Sobel edge operator.

Sobel edge operator estimate gradient in x-direction with the help of the gradient component called $G_x$ on selected image [1]. Sobel edge operator estimate gradient in y-direction with the help of gradient component called $G_y$ on selected image. The gradient components $G_x$ and $G_y$ are shown in figure 2.

| -1 | 0 | +1 |
|----|---|----|
| -2 | 0 | +2 |
| -1 | 0 | +1 |

$G_x$

| +1 | +2 | +1 |
|----|----|----|
| 0  | 0  | 0  |
| -1 | -2 | -1 |

$G_y$

Figure 2. Gradient component masks in x-direction and y- direction

The absolute gradient magnitude or edge strength is calculated by using following formula-

$$|G| = |G_x| + |G_y| \tag{4}$$

*Step 3:* The direction of the edge is computed using the incline in the x and y directions. Whenever the value of $G_x$ is equal to zero, the edge direction has to be equal to 90 degrees or 0 degrees, depending on what the value of $G_y$. If $G_y$ has a value of zero, the edge direction will be equal to 0 degrees. Otherwise the edge direction will be equal to 90 degrees.

The equation for finding the edge directions of an image is-

$$\theta = -\tan^{-1}(G_y/G_x) \tag{5}$$

*Step 4:* Relate the edge direction to the direction of the image that can be traced. The four directions that can be formed near to neighboring pixels are horizontal (0 Degrees), along the plus diagonal (45 Degrees), perpendicular direction (90 Degrees) and along the minus diagonal (135 Degrees).
*Step 5:* After knowing the edge directions, the edge thinning technique called non-maximum suppression is applied. Non-maximum suppression is used to trace the edge along the boundary of edges and conquer any pixel value that cannot be considered as part of an edge. This will give thin line in an output image.
*Step 6:* Finally, hysteresis is being used to remove streaks. Streaking is the breaking up of an edge outline that is caused due to fluctuation of operating above and below the threshold. For hysteresis process requires two levels of threshold that are top level and bottom level threshold.

*Advantages:*
- Finding of errors is effective and easy with the help of probability.
- Better Localization of edges and response for edges.
- Improving signal to noise ratio, as non-maximum suppression is used.
- Better edge detection in the noise state with the help of Thresholding method.

*Disadvantages:*
- Complex Computations.
- Time consuming.

## 6. LSBMR Algorithm

The Least Significant Bit Matching Revisited (LSBMR) [5] algorithm was used for gray-scaled cover images. The algorithm uses two pixels of the cover image as embedding unit to hide out the secret message. From the two pixels, the first pixel $x_i$ is used to conceal the secret message bit $m_i$ and the binary relationship between the pixels ($x_i$, $x_{i+1}$) value is used to hide out another message bit $m_{i+1}$. The relationship between both pixels is calculated with the help of floor function. The floor function is shown as follows:

$$f(x_i, x_{i+1}) = LSB(floor\left(\frac{x_i}{2}\right) + x_{i+1}) \qquad (6)$$

LSBMR eliminates the asymmetry property caused by the basic LSB approach. It resists the Steganalytic attacks that exploit the asymmetry property of the Stego-image. As it uses add one or subtract one schema, the probability of the expected number of modifications reduced from 0.5 in case of LSB to 0.375 for LSBMR for the same payload capacity. Hence, the statistical detectability of the image steganography is low. LSBMR performs data embedding on pixel pairs using following four cases:

*Case 1*: LSB ($x_i$) = $m_i$ & f ($x_i$,$x_{i+1}$) = $m_{i+1}$
  $(x'_i, x'_{i+1}) = (x_i, x_{i+1})$ ;

*Case 2:* LSB ($x_i$) = $m_i$ & f ($x_i$,$x_{i+1}$) ≠ $m_{i+1}$
  $(x'_i, x'_{i+1}) = (x_i, x_{i+1} \pm 1)$ ;

*Case 3:* LSB ($x_i$) ≠ $m_i$ & f ($x_i$,$x_{i+1}$) = $m_{i+1}$
  $(x'_i, x'_{i+1}) = (x_i - 1, x_{i+1})$ ;

*Case 4:* LSB ($x_i$) ≠ $m_i$ & f ($x_i$,$x_{i+1}$) ≠ $m_{i+1}$
  $(x'_i, x'_{i+1}) = (x_i + 1, x_{i+1})$ ;

Where $m_i$ and $m_{i+1}$ are message bits, $x_i$ and $x_{i+1}$ is a pixel pair before data embedding and $x'_i$ & $x'_{i+1}$ is a pixel pair after data embedding. The embedding cannot be performed for pure pixels that have either a negligible or supreme allowable value. The message embedding rate is discovered by the pseudo-random sequence generator procedure.

## 7. Proposed Mechanism

The proposed scheme of edge adaptive steganography is based on the Threshold value and LSBMR algorithm. The threshold value is the intensity difference between two consecutive pixels; these two pixels are considered as an embedding unit for LSBMR. LSBMR takes message bits and pixel pair as input for data encoding. The detailed data embedding algorithm are as follows.

### 7.1   Data Embedding

Following figure 3 shows the block diagram of proposed technique used for data embedding.

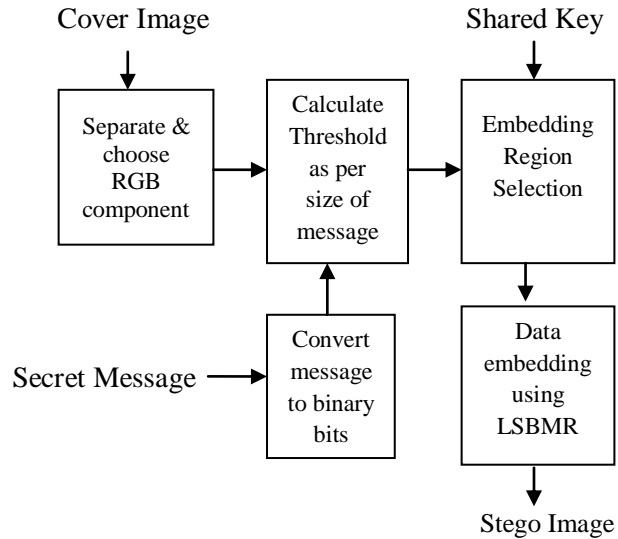

Figure 3. Data Embedding

The proposed data embedding technique are described in steps as follows:

*Step 1:* Separate RGB components of the cover image and choose the desired component (R, G, and B) for data embedding.

*Step 2:* Convert the input message into binary bits with the help of ASCII value conversion.

*Step 3:* Evaluate the threshold value according to the size of secret message. The threshold value is the intensity value difference between two pixels having difference greater or equal to the threshold value.

*Step 4:* Select the region, according to the threshold value for generating embedding pairs. Save the encoding pairs as secret key.

*Step 5:* The LSBMR algorithm is used for embedding the secret message bits into the cover image.

*Step 6:* Rewrite the changes into cover image occurred during data embedding.

*Step 7:* Combine the RGB components of the cover image and save the image as Stego image.

### 7.2 Data Extraction

Following figure 4 shows the block diagram of proposed technique used for data extraction.

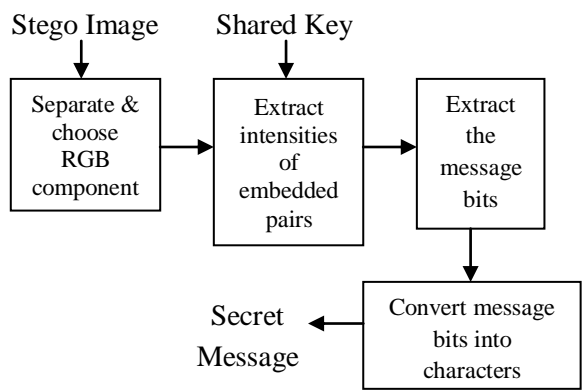

Figure 4. Data Extraction

The proposed data extraction technique is described in steps as follows:

*Step 1:* Separate the RGB components of Stego image and select the component from R, G, and B that was used during data embedding process.

*Step 2:* Extract the intensities of encoding pairs using a shared key.

*Step 3:* Extract the message bits $m_i$ and $m_{i+1}$ from intensities that are calculated in step 2 and using LSBMR algorithm.

*Step 4:* Convert binary bits of messages in ASCII value and then convert the ASCII value into its character to get the original message.

## 8. Experimental Results and Discussion

The tool used for implementing the work is MATLAB 7.6 and above versions. The objective of the work is to extend the use of edge adaptive steganographic techniques for color images with some modifications to the parameters used before and analyze the performance of the proposed method. The proposed method is implemented on nearly 100 color images. The performance of the proposed method is evaluated and compared on the basis of two parameters and the parameters are: Mean Square Error (MSE) and Peak Signal to Noise Ratio (PSNR) and they are computed as follows:

### 8.1  Mean Square Error

The mean square error (MSE) [1] is a statistical measure of how far modified values are different from the actual values. It is mostly used in time series, but can be used in any sort of statistical estimate [3]. Here in this project, it could be applied to pixel pairs, where one set is "Cover Image" and the other is a "Stego Image".

Steps for Calculating Mean Square Error:

1) Take Cover Image (CI) and Stego Image (SI) as inputs.
2) Subtract Stego Image values (pixel pair values) from Cover Image values.
3) Take the absolute value of each row. That is, if the difference is negative, remove the negative sign. If it is positive, keep it as is.
4) Add up the absolute values of each row.
5) Take the square of the resultant.
6) Normalize the total number of rows and columns.ie. m x n.

$$\text{MSE} = \frac{1}{m \times n} \sum_{i=1}^{m-1} \sum_{j=1}^{n-1} (CI - SI)^2 \qquad (7)$$

In above equation, CI= Cover Image, SI= Stego Image
In simple words MSE indicates average amount of modifications to the pixels.

### 8.2  Peak Signal-to-Noise Ratio

The peak signal-to-noise ratio (PSNR) [1] is the ratio between a signal's maximum power and the power of the signal's noise. Here the PSNR is used to measure the quality of Stego image. Each pixel of an image has a color value that may change after an image gets modified. Signals can have a broad ever changing range; therefore PSNR is normally expressed in decibels [3]. Decibel representation is in logarithmic scale. Peak Signal to Noise Ratio is computed using the formula:

$$PSNR = 10 \times log_{10} \left(\frac{256 \times 256}{MSE}\right) \qquad (8)$$

The PSNR value indicates the quality of Stego image after modification of Cover image.

The MSE and PSNR parameters are calculated for several Cover images. Some of the Cover images are taken as example for result calculation and comparison with existing techniques. Following figure 5 considered as an example for the experiment.

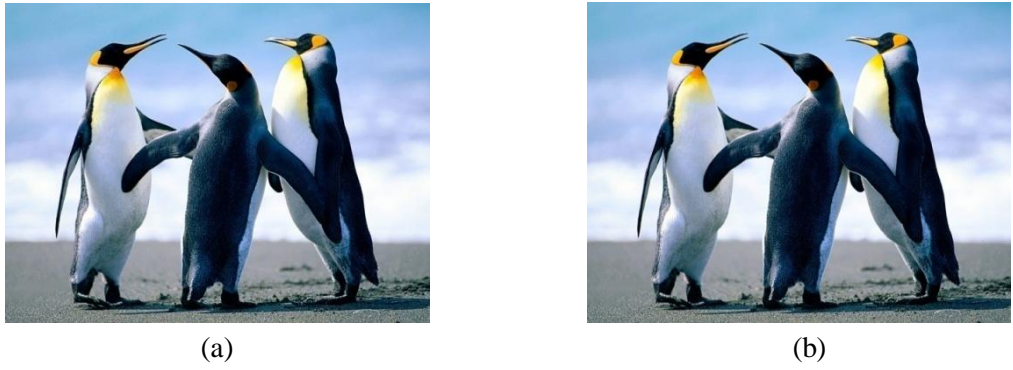

(a)                  (b)

Figure 5. (a) Cover Image and (b) Stego Image

The results of MSE and PSNR for Cover Image figure 5 (a) and Stego Image figure 5 (b) using different message lengths are shown in Table 1.

Table 1. MSE and PSNR values

| Message Length approximately (Bits) | MSE Values | | | PSNR Values | | |
|---|---|---|---|---|---|---|
| | Sobel | Canny | Threshold Value | Sobel | Canny | Threshold Value |
| 400 | 0.000026 | 0.000026 | 0.000025 | 93.968707 | 93.968707 | 94.111111 |
| 600 | 0.000031 | 0.000031 | 0.000028 | 93.319299 | 93.319299 | 93.631876 |
| 900 | 0.000050 | 0.000050 | 0.000047 | 91.210765 | 91.210765 | 91.439394 |
| 1200 | 0.000060 | 0.000060 | 0.000054 | 90.400432 | 90.400432 | 90.820524 |

The table 1 shows four parameters for analyzing the results. First parameter is message length, which is the length of data bits to be embedded inside the Cover image. Second parameter is an edge detection technique for detecting edges of an image with the help of Sobel and Canny edge detection methodology, and then LSBMR is implemented in order to embed the detected edges. Third parameter is Mean Square Error for measuring the modification rate. The fourth parameter is Peak Signal-to-Noise Ratio for measuring the quality of an image.

Threshold Values and the length of messages for figures 5 (a) and 5 (b) are shown in Table 2.

Table 2. Message length and Threshold value

| Message length approximately (Bits) | Threshold Value |
|---|---|
| 400 | 170 |
| 600 | 164 |
| 900 | 151 |
| 1200 | 145 |

The table 1 shows improved values of PSNR and MSE parameters. From the results we can say that the proposed Threshold Value mechanism is giving better results than Sobel and Canny edge detection technique. The proposed approach achieves better quality of an image and reduces the modification rate as compared to Sobel and Canny edge detection technique. The table 2 shows the variation between threshold value and length of the message. The length of the message is inversely proportional to a threshold value.

Another example is considered for evaluation of results to show the variations in the PSNR and MSE values. Figure 6 consists of the cover image and stego image.

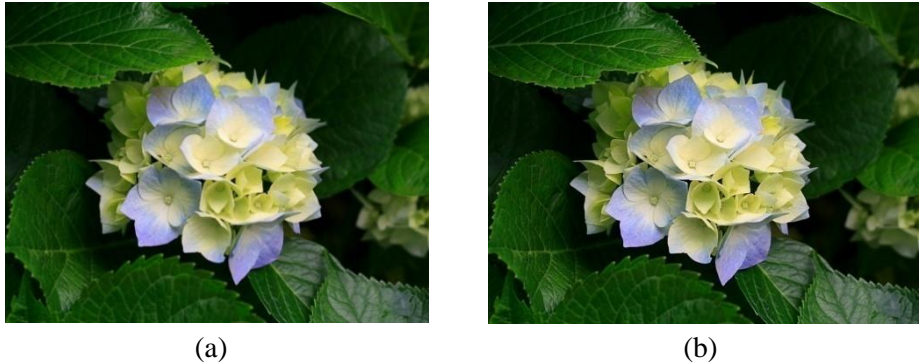

(a)                 (b)

Figure 6. (a) Cover Image and (b) Stego Image

The results of evaluation of MSE and PSNR values on Cover Image i.e. fig. 6 (a) and Stego Image i.e. fig. 6 (b) using different message lengths is shown in Table 3.

Table 3. MSE and PSNR values

| Message Length approxi. (Bits) | MSE Values | | | PSNR Values | | |
|---|---|---|---|---|---|---|
| | Sobel | Canny | Threshold Value | Sobel | Canny | Threshold Value |
| 400 | 0.000028 | 0.000028 | 0.000024 | 93.631876 | 93.631876 | 94.333875 |
| 600 | 0.000030 | 0.000030 | 0.000025 | 93.441643 | 93.441643 | 94.184104 |
| 900 | 0.000048 | 0.000048 | 0.000047 | 91.361839 | 91.361839 | 91.380765 |
| 1200 | 0.000064 | 0.000060 | 0.000059 | 90.102854 | 90.102854 | 90.431343 |

Threshold Values and the length of messages for figures 6 (a) and 6 (b) are shown in Table 4.

Table 4. Message length and Threshold value

| Message length approximately (Bits) | Threshold Value |
|---|---|
| 400 | 119 |
| 600 | 113 |
| 900 | 101 |
| 1200 | 95 |

The table 3 shows improved values of PSNR and MSE parameters. From the results we can say that the proposed Threshold Value technique is giving better results than Sobel and Canny edge detection technique. The proposed technique provides better quality of an image as well as the modification rate of an image gets reduced as compared to Sobel and Canny edge detection technique. The table 4 shows the variation in Threshold Value in comparison with the length of a message. The length of the message is inversely proportional to a threshold value. That means, as the length of the message increases the threshold value gets decreases.

## 9. Conclusion and Future Scope

First the quality of an image and modification rate for an image is estimated using Sobel and Canny edge detection techniques. Then the quality of an image and modification rate for an image is estimated using proposed Threshold Value approach. The quality of an image is measured by parameter PSNR and Modification rate for an image is measured by parameter MSE for all the mentioned methods. The conclusions are carried out by considering the image in figure 5.

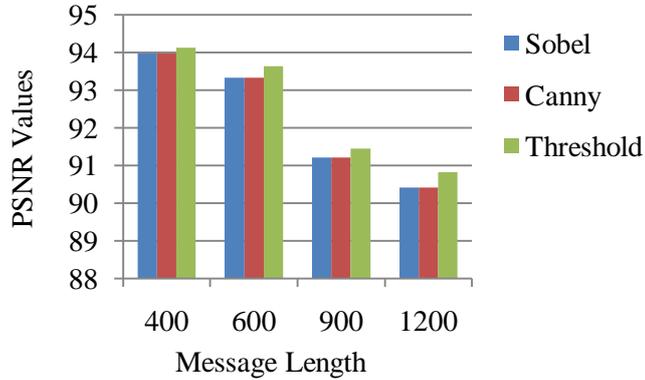

Figure 7. PSNR values of an Image

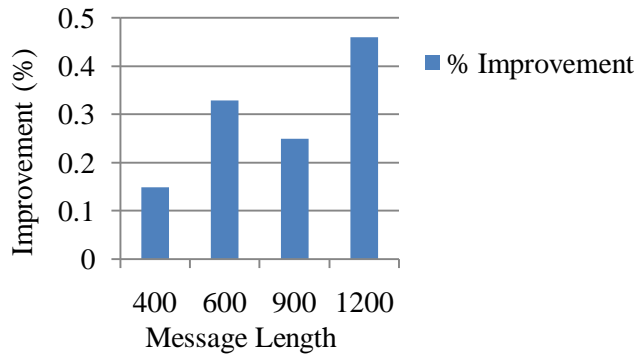

Figure 8. Percentage improvement in the quality of an image

The proposed Threshold Value approach results achieve about 0.2 to 0.6 % of improvement in the quality of an image as compared to Sobel and Canny based technique. The percentage improvement results in the quality of an image are shown in figure 8.

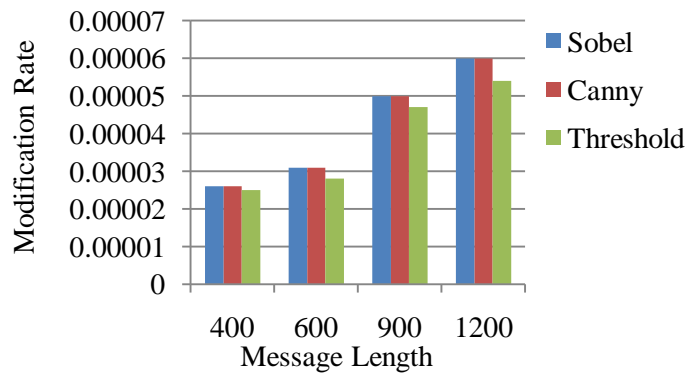

Figure 9. MSE values of an Image

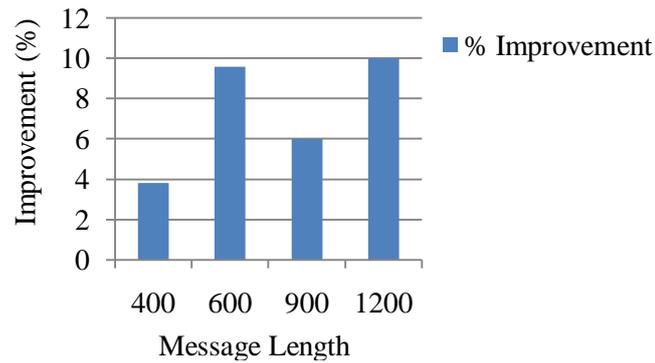

Figure 10. Percentage improvement in the modification rate of an image

Similarly, the proposed Threshold Value approach achieves 4 to 10 % of improvement in the modification rate of an image as compared to Sobel and Canny based approach. The percentage improvement results in the modification rate of an image are shown in figure 10.

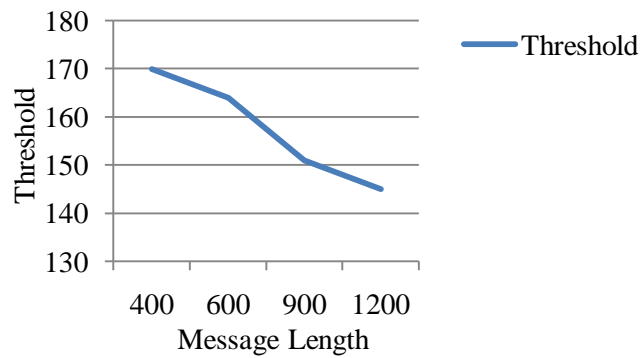

Figure 11. Threshold Value Vs Message Length

Threshold Value is inversely proportional to the length of the message. That means, as the length of the message increases the threshold value decreases. It helps in embedding number of data bits during data embedding process. Figure 11 shows results variation between Threshold Value and Message length.

The proposed Threshold Value approach can be implemented in the future for following real world applications:

1) The proposed approach can be implemented to other cover medium such as audio/video.
2) The proposed approach can be applied on gray scale images.
3) The proposed approach can be used for frequency domain, type of data embedding.
4) The proposed approach can be implemented for real time Image Authentication System.

# References

[1] Zohreh Fouroozesh and Jihad Al jaam, (2014) "Image Steganography based on LSBMR using Sobel Edge Detection", *IEEE*.

[2] B. Sharmila and R. Shanthakumari, (2012) "Efficient Adaptive Steganography for color images based on LSBMR algorithm", *ICTACT Journal on image and video processing*, volume: 02, issue: 03.

[3] Weiqi Luo, Fangjun Huang and Jiwu Huang, (2010) "Edge Adaptive Image Steganography Based on LSB Matching Revisited", *IEEE Transactions on Information Forensics and Security*, vol. 5, no. 2.

[4] Nidhi Grover and A. K. Mohapatra, (2013) "Digital Image Authentication Model Based on Edge Adaptive Steganography", *Second International Conference on Advanced Computing, Networking and Security.*

[5] J. Mielikainen, Jarno. (2006) "LSB matching revisited", *Signal Processing Letters, IEEE* 13, no. 5, 285-287.

[6] Bill Green, Canny Edge Detection Tutorial, 2002.

[7] Mamta Juneja, Parvinder Singh Sandhu, (2009) "Performance Evaluation of Edge Detection Techniques for Images in Spatial Domain", *International Journal of Computer Theory and Engineering,* Vol. 1, No. 5.

[8] Syed Sameer Rashid, Swati R. Dixit and A. Y. Deshmukh, (2014) "VHDL Based Canny Edge Detection Algorithm", *International Journal of Current Engineering and Technology*, Vol.4, No.2.

[9] Nitin Jain, Sachin Meshram, Shikha Dubey, (2012) "Image Steganography Using LSB and Edge Detection Technique", *International Journal of Soft Computing and Engineering (IJSCE)* ISSN: 2231-2307, Volume-2, Issue-3.

[10] Yang C. H, Weng C. Y, Wang S. J, and Sun H. M , (2008) "Adaptive data hiding in edge areas of images with spatial LSB domain systems", *Transaction on Information Forensics Security*, Vol. 3, No. 3, pp. 488–497.

[11] Fridrich J., Goljan J., and Du R., (2001) "Detecting LSB steganography in color, and gray-scale images", *IEEE Multimedia*, Vol. 8, No. 4, pp. 22–28.

[12] Deepali Singla and Mamta Juneja, (2014) "An Analysis of Edge Based Image Steganography Techniques in Spatial Domain", *Proceedings of 2014 RAECS UIET Punjab University Chandigarh*, 06 – 08.

[13] Zhenhao Zhu, Tao Zhang, and Baoji Wan, (2013) "A Special Detector for the Edge Adaptive Image Steganography Based on LSB Matching Revisited", *10th IEEE International Conference on Control and Automation (ICCA) Hangzhou, China*, June 12-14.